\documentclass[
 amsmath,amssymb,
 aps,
 pra,
 twocolumn,
]{revtex4-2}

\usepackage{graphicx}
\usepackage{dcolumn}
\usepackage{bm}
\usepackage{multirow}
\usepackage{array}

\usepackage{times}
\usepackage{amsmath}
\usepackage{amssymb}
\usepackage{braket}

\newcounter{lastnote}

\usepackage{hyperref}
\usepackage{xcolor}
\hypersetup{
	colorlinks   = true, 
	urlcolor     = blue, 
	linkcolor    = blue, 
	citecolor   = blue 
}

\begin{document}
\preprint{APS/123-QED}

\title{Near-ultrastrong nonlinear light-matter coupling in superconducting circuits} 
\author{Yufeng Ye$^{1,2}$, Jeremy B. Kline$^{1,2}$, Alec Yen$^{1,2}$, Gregory Cunningham$^{2,3}$, Max Tan$^{1,2}$, Alicia Zang$^{1,2}$, Michael Gingras$^{4}$, Bethany M. Niedzielski$^{4}$, Hannah Stickler$^{4}$, Kyle Serniak$^{2,4}$, Mollie E. Schwartz$^{4}$, Kevin P. O'Brien$^{1,2}$}

\email[Correspondence email address: ]{kpobrien@mit.edu}%

\affiliation{$^{1}$Department of Electrical Engineering and Computer Science, Massachusetts Institute of Technology, Cambridge, MA 02139, USA,}
\affiliation{$^{2}$Research Laboratory of Electronics, Massachusetts Institute of Technology, Cambridge, MA 02139, USA,}
\affiliation{$^{3}$Harvard John A. Paulson School of Engineering and Applied Sciences, Harvard University,
Cambridge, MA 02138, USA,}
\affiliation{$^{4}$Lincoln Laboratory, Massachusetts Institute of Technology, Lexington, MA 02421, USA}

\date{\today} 

\begin{abstract}
The interaction between an atom and an electromagnetic mode of a resonator is of both fundamental interest and is ubiquitous in quantum technologies. Most prior work studies a linear light-matter coupling of the form $g \hat{\sigma}_x (\hat{a} + \hat{a}^\dagger)$, where $g$ measured relative to photonic ($\omega_a$) and atomic ($\omega_b$) mode frequencies can reach the ultrastrong regime ($g/\omega_{a}\!>\!10^{-1}$). In contrast, a nonlinear light-matter coupling of the form $\frac{\chi}{2} \hat{\sigma}_z \hat{a}^\dagger \hat{a}$ has the advantage of commuting with the atomic $\hat{\sigma}_z$ and photonic $\hat{a}^\dagger\hat{a}$ Hamiltonian, allowing for fundamental operations such as quantum-non-demolition measurement. However, due to the perturbative nature of nonlinear coupling, the state-of-the-art $\chi/\text{max}(\omega_a, \omega_b)$ is limited to $\!<\!10^{-2}$. Here, we use a superconducting circuit architecture featuring a quarton coupler to experimentally demonstrate, for the first time, a near-ultrastrong $\chi/\text{max}(\omega_a, \omega_b)= (4.852\pm0.006)\times10^{-2}$ nonlinear coupling of a superconducting artificial atom and a nearly-linear resonator. We also show signatures of light-light nonlinear coupling ($\chi\hat{a}^\dagger\hat{a}\hat{b}^\dagger\hat{b}$), and $\chi/2\pi = 580.3 \pm 0.4 $ MHz matter-matter nonlinear coupling ($\frac{\chi}{4}\hat{\sigma}_{z,a}\hat{\sigma}_{z,b}$) which represents the largest reported $ZZ$ interaction between two coherent qubits.
Such advances in the nonlinear coupling strength of light, matter modes enable new physical regimes and could lead to applications such as orders of magnitude faster qubit readout and gates. 

\end{abstract}

\maketitle

\section{Introduction}
Linear light-matter coupling can be modeled by the quantum Rabi Hamiltonian \cite{rabi1937space} ($\hbar=1$ hereafter),
\begin{equation}
\hat{H}_\text{linear} = \omega_a \hat{a}^\dagger \hat{a} + \frac{\omega_b}{2} \hat{\sigma}_z + g (\hat{a}^\dagger + \hat{a} ) \hat{\sigma}_x,
\label{eq:H_Rabi}
\end{equation}
where $\hat{a}$ is the annihilation operator of a resonator serving as the photonic or light-like mode and $\hat{\sigma}_{x,z}$ are Pauli operators of a two-level system or qubit serving as the atomic or matter-like mode. Eq.~(\ref{eq:H_Rabi}) describes \textit{linear} light-matter coupling in the language of nonlinear optics \cite{Boyd} because the interaction $g (\hat{a}^\dagger + \hat{a} ) \hat{\sigma}_x$ is linear in $(\hat{a}^\dagger + \hat{a} )$ and $\hat{\sigma}_x$, which represent the electric field of the light and matter dipole, respectively. 
In general, the coupling strength is important both for demonstrating fundamental physics and practical quantum technologies where stronger coupling fundamentally enables faster entanglement operations \cite{MargolusLevitin} that are less limited by qubit or photon decoherence times. 
Across physical platforms such as atoms, molecules, and superconducting circuits, the magnitude of $g$ can exceed the decay rates of the system in a regime known as strong coupling \cite{ultrastrong_review, USC_RMP}. Furthermore, relative to the frequency of the photonic mode $\omega_{a}$ (assumed to be near-resonant with the atomic mode $\omega_{a}\approx \omega_b$), the normalized coupling $\eta := g/\omega_{a}$ \cite{ultrastrong_review, USC_RMP} has reached the ultrastrong ($\eta\!>\!10^{-1}$) \cite{firstUSC} and the deep-strong regime ($\eta\!>\!1$) \cite{first_DSC}. 
\begin{figure*}
    \includegraphics[width=1\textwidth]{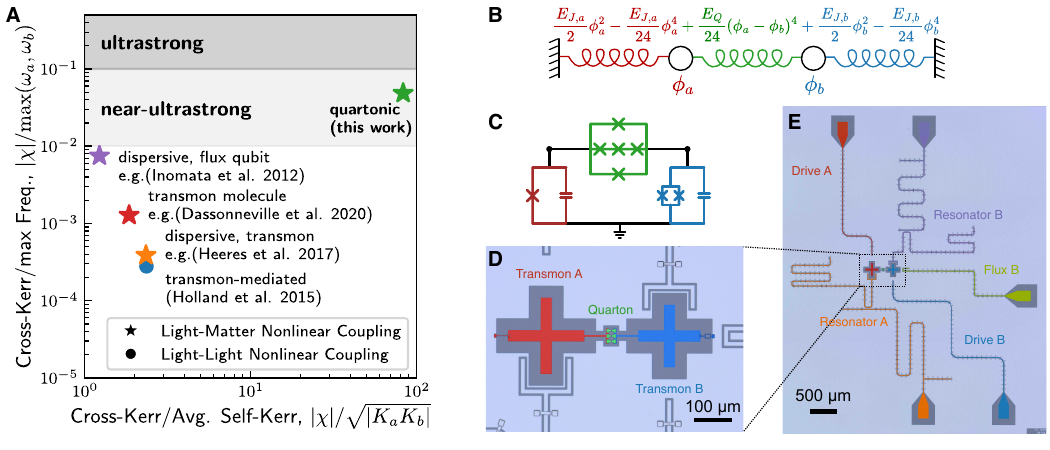}
    \caption{\textbf{Quarton coupler physics and experimental device.} (\textbf{A}) Parameter landscape of light-matter nonlinear coupling (4-wave-mixing Kerr effect). The presented quartonic scheme reaches near-ultrastrong nonlinear coupling with normalized nonlinear coupling (cross-Kerr normalized by frequency) of $\tilde{\eta} = (4.852\pm0.006)\times10^{-2}$, which is achieved without a trade-off of larger self-Kerr ($|\chi|/\sqrt{|K_a K_b|} = 83.2$). (\textbf{B}) Spring-mass analogue of the quarton coupler device circuit, colored equations show the potential energies up to $\phi^4$. (\textbf{C}) Effective circuit of the quarton-coupled two transmon device.  (\textbf{D}) False-colored micrograph of transmon A (red), B (blue) coupled by a gradiometric quarton coupler (green). (\textbf{E}) False-colored micrograph of entire chip, including a flux-bias line for transmon B (yellow), drive lines for transmon A (red), B (blue), and Purcell-protected readout resonators A (orange), B (purple).} 
    \label{fig: exp_1}
\end{figure*}

However, large $g$ significantly alters the photonic and atomic eigenstates because the linear coupling $g \hat{\sigma}_x (\hat{a} + \hat{a}^\dagger)$ does not commute with the atomic $\hat{\sigma}_z$ and photonic $\hat{a}^\dagger\hat{a}$ Hamiltonian. This fundamentally precludes quantum-non-demolition (QND) measurements of either photon number or qubit state, and leads to effects such as unwanted Purcell decay \cite{PurcellDecay} or errors in quantum operations when $g$ cannot be precisely switched off \cite{residualZZ}. 
As such, in important applications such as QND qubit readout \cite{wallraff2005dispersive, dispersive_spinqubit}, QND single photon detection \cite{number_splitting, cavityQED-numbersplit}, and certain single \cite{krastanov2015} or two-qubit gates \cite{RIPgate_experiment, cat_exponoff}, the cross-Kerr interaction is used instead. 
This is a type of \textit{nonlinear} light-matter coupling:
\begin{equation}
\hat{H}_{\text{nonlinear}} = \omega_a \hat{a}^\dagger \hat{a} + \frac{\omega_b}{2} \hat{\sigma}_z + \frac{\chi}{2}\hat{a}^\dagger \hat{a} \hat{\sigma}_z,
\label{eq:H_crossKerr}
\end{equation}
where (unlike Eq.~(\ref{eq:H_Rabi})) the coupling is nonlinear in the electric field since $\hat{a}^\dagger \hat{a}$ and $\hat{\sigma}_z$ represent the electromagnetic energy ($\sim$ field squared) of the light and matter mode, respectively. 
Since most nonlinear effects in nature tend to be a higher order process \cite{Boyd}, the nonlinear $\chi$ is usually perturbatively small compared to the linear $g$. In fact, the state-of-the-art approach for realizing Eq.~(\ref{eq:H_crossKerr}), is to simply use the linear coupling of Eq.~(\ref{eq:H_Rabi}) in the dispersive regime \cite{blais2004} ($g\!\ll\!\Delta = |\omega_a - \omega_b|$) which produces an effective cross-Kerr $\chi = g^2 / \Delta \ll g$. 
Analogous to strong linear coupling, the strong nonlinear coupling regime is defined as $\chi$ greater than decay rates, and has been demonstrated in physical platforms such as atoms \cite{first-strong-disp} and superconducting circuits \cite{number_splitting}, leading to observation of phenomenon such as photon-number splitting of the qubit transition \cite{number_splitting}. The largest reported cross-Kerr uses dispersive coupling with a superconducting flux qubit for $\chi/2\pi = 80 $ MHz \cite{Inomata2012}.
To better characterize the strength of the nonlinear coupling beyond strong coupling, we hereby define the normalized \textit{nonlinear} coupling  $\tilde{\eta} := \chi/\text{max}(\omega_a,\omega_b)$, analogous to the normalized \textit{linear} coupling $\eta := g/\omega_{a}$ \cite{ultrastrong_review, USC_RMP}. Note that $\tilde{\eta}$ uses a more representative definition because $\chi$ measurements do not require $\omega_a \approx \omega_b$, in contrast to avoided-crossing measurements for $g$. Natural definitions of ultrastrong ($\tilde{\eta}\!>\!10^{-1}$) and deep-strong ($\tilde{\eta}\!>\!1$) nonlinear coupling then follow. To the best of our knowledge, the largest reported $\tilde{\eta}$ is only $6\times10^{-3}$ \cite{Inomata2012}, which is close to two orders of magnitude away from the ultrastrong nonlinear coupling regime ($10^{-1}$). This motivates us to introduce an intermediate regime, near-ultrastrong nonlinear coupling, to cover the range of $10^{-2}\!<\!\tilde{\eta}\!<\!10^{-1}$ that bridges existing capabilities and the ultrastrong regime. 
More than two decades after the first demonstration of strong nonlinear coupling \cite{first-strong-disp}, it is of fundamental interest to now explore the subsequent regimes of near-ultrastrong, ultrastrong, and deep-strong nonlinear coupling. 
Furthermore, it is also of great practical interest since cross-Kerr-based applications have operation times $\propto\!1/\chi$ \cite{krastanov2015, walter2017} or even $\propto\!1/\chi^2$ \cite{ShrutiRIPgate}, therefore realizing orders of magnitude stronger $\chi$ can lead to proportionally significant improvements including ultrafast qubit readout \cite{Quartonic_Readout}, photon detection \cite{Arne_SPD}, and qubit gates \cite{krastanov2015, RIPgate_experiment}.

Here, we show the first experimental demonstration of the near-ultrastrong nonlinear light-matter coupling regime in any physical platform. We demonstrate, in superconducting circuits, a normalized nonlinear coupling of $\tilde{\eta} = (4.852\pm0.006)\times10^{-2}$. Using a gradiometric quarton coupler, at one flux-bias point, we measure cross-Kerr coupling of $\chi/2\pi = 366.0 \pm 0.5 $ MHz between two transmons, with one serving as a nonlinear artificial atom and the other as a nearly-linear resonator with self-Kerr anharmonicity of only $0.76 \pm 0.08$ MHz. Crucially, the large nonlinear coupling $\chi$ is not a result of large self-nonlinearity (self-Kerrs $K_a, K_b$), as our system exhibits $\chi/\sqrt{K_a K_b} > 80$, whereas to the best of our knowledge, all previous light-matter nonlinear couplings have been restricted to $\chi/\sqrt{K_a K_b} \sim O(1)$ (see Fig.~\ref{fig: exp_1}A). This allows us to simulate the regime of light-light nonlinear coupling \cite{Holland2015} at the same $\chi$, observing photon-number splitting \cite{number_splitting} of the transitions of both transmons. We then flux-bias the quarton coupler for maximal nonlinear coupling between two nonlinear qubit modes, where we measure a matter-matter nonlinear coupling of $\chi/2\pi = 580.3 \pm 0.4 $ MHz, which is (to the best of our knowledge) the largest reported cross-Kerr ($ZZ$) coupling between two coherent qubits.

In the following sections, we begin by presenting the device and the key operating principles based on our previous theory proposal \cite{quartonPRL}, then we present spectroscopy results showing transmon self-Kerr tuning with quarton coupler flux bias which enables linearization (self-Kerr near zero) of one transmon at a flux-bias point. At said point, we measure cross-Kerr coupling strength with spectroscopy, showing near-ultrastrong nonlinear light-matter coupling. Finally, we modify the spectroscopy to simulate light-light nonlinear coupling, and demonstrate matter-matter nonlinear coupling at another quarton coupler flux-bias point.

\section{Results}
\subsection{Quarton coupler circuit}
Superconducting circuits is a leading platform for the study and control of light-matter interaction \cite{blais2021,gu2017review}. By exploiting the nonlinear kinetic inductance of the Josephson junction (JJ) to make quantum oscillators with nonlinear energy levels, high coherence artificial atoms or qubits can be realized. 
We use here a common type of superconducting qubit, known as the transmon \cite{KochTransmon}, which can be understood as a microwave resonator with added self-Kerr nonlinearity ($K<0$) from the JJ:
\begin{equation}
\hat{H}_{\text{transmon}} = \omega_b \hat{b}^\dagger \hat{b} + \frac{K}{2} \hat{b}^{\dagger2} \hat{b}^2 + \dots \approx \frac{\omega_b}{2} \hat{\sigma}_z,
\label{eq:H_transmon}
\end{equation}
The key insight is that since adding a self-Kerr of $K$ turns a linear resonator (photonic) mode into a qubit (atomic) mode, then removing $K$ linearizes a transmon qubit into a resonator. This is achieved using the quarton coupler we proposed in \cite{quartonPRL}, which can induce an opposite-signed (positive) self-Kerr to transmons while facilitating large cross-Kerr between them. This ``quartonic'' approach allows us to achieve large cross-Kerr $\chi$ without causing a large self-Kerr $K$ that would otherwise compromise the linearity of the photon mode.
We contrast our approach with the state-of-the-art in Fig.~\ref{fig: exp_1}A which shows the parameter landscape of light-matter nonlinear coupling (including 1 additional case of light-light nonlinear coupling \cite{Holland2015}) with 4-wave-mixing Kerr effect, wherein we use calculated $K_a$ when not provided \cite{Inomata2012, Dassonneville2020}. 
To the best of our knowledge, all previous experimental cross-Kerr demonstrations \cite{Inomata2012, Dassonneville2020, heeres2017, Holland2015} 
are limited to $|\chi|/\text{max}(\omega_a,\omega_b) < O(10^{-2})$ and a trade-off appears where larger nonlinear coupling is accompanied by disproportionately larger self-nonlinearity (decreasing $|\chi|/\sqrt{|K_a K_b|}$). Existing demonstrations are also limited to $|\chi|/\sqrt{|K_a K_b|} \sim O(1)$, as expected when cross-Kerr interactions are dominated by first-order effects which satisfy $|\chi|/\sqrt{|K_a K_b|} = 2$ \cite{BBQ}.

The circuit realized in this work, shown in Fig.~\ref{fig: exp_1}C, consists of two transmons (red, blue) galvanically coupled by a gradiometric quarton coupler (green). The gradiometric circuit topology is inspired by other works \cite{ATS, G-SNAIL}.  The  circuit's potential energy can be written in terms of Josephson energies $E_J$ and the node superconducting phases $\phi_a$, $\phi_b$:
%
%
\begin{equation}
\label{eq:potential}
\begin{aligned}
U & =-E_{Ja}\cos{(\frac{\tilde{\phi}_s}{2})}\cos(\phi_a)-E_{Jb}\cos(\phi_b) \\
& -3E_J\cos(\frac{\phi_a-\phi_b}{3}) \\
& - \alpha E_J\cos(\tilde{\phi}_{q\Sigma})\cos(\phi_a-\phi_b),
\end{aligned}
\end{equation}
%
%
%
%
where we have assumed that the two nominally identical loops of the gradiometric quarton are identically flux-biased. The gradiometric quarton then behaves as a quarton with $\tilde{\phi}_{q\Sigma}$ flux tunable $\alpha$, which varies its ratio of linear coupling, $(\phi_a - \phi_b)^2$, to nonlinear coupling, $(\phi_a - \phi_b)^4$ \cite{quartonPRL}. At $\alpha \cos{(\tilde{\phi}_{q\Sigma})} = -1/3$, the quarton coupling potential $\frac{E_Q}{24}(\phi_a-\phi_b)^4 + \dots$ is to leading order quartic with effective Josephson energy $E_Q = \frac{8}{27} E_J$.
 
 The behavior of this circuit can be understood with a spring-mass analogue as shown in  Fig.~\ref{fig: exp_1}B, where we treat the two node phases $\phi_a$, $\phi_b$ as position coordinates, and the transmon JJs act as slightly nonlinear springs with spring constant $E_J$. Keeping terms up to $O(\phi^4)$, the quarton acts as a purely nonlinear coupling spring with potential energy $\frac{E_{Q}}{24}(\phi_a - \phi_b)^4$. This allows cancellation of the $-\frac{E_{Ja}}{24}\phi_a^4$ self-nonlinearity of the $\phi_a$ mode (if $E_Q \approx E_{Ja}$) while creating a strong $\phi_a^2\phi_b^2$ nonlinear coupling between the two modes. Writing the $\phi$ operators in the Fock basis, one can see that this $\phi_a^2\phi_b^2$ coupling leads to a non-perturbative cross-Kerr term $\propto \hat{a}^\dagger\hat{a}\hat{b}^\dagger\hat{b}$. 

A false-colored micrograph of our  device is shown in Fig.~\ref{fig: exp_1}E, with a close-up of the two transmons in Fig.~\ref{fig: exp_1}D. Transmon A, on the left, will be linearized into a light-like mode with near zero self-Kerr anharmonicity, while transmon B, on the right, will remain a nonlinear qubit or matter-like mode. Both transmons have drive lines and Purcell-protected \cite{PurcellFilter2014} readout resonators labeled A and B, which are capacitively coupled to transmons A and B, respectively. The chip also includes a local flux-bias line to tune the SQUID in transmon B, and the chip package has a global coil to bias the gradiometric quarton coupler. See Appendix \ref{sec:methods} for more details about the experimental setup.

In addition to the quarton and SQUID loops, the upper and lower ground plane around the circuit form two loops with the JJs of the circuit (see Fig.~\ref{fig: exp_1}D). Symmetric flux in these loops produces an unimportant screening current in the ground plane, while asymmetric flux in these loops ($\tilde{\phi}_{g\Delta}$) will bias the junctions. 
We calibrate the local and global flux bias such that  $\tilde{\phi}_{g\Delta}\approx0$, so that only the SQUID ($\tilde{\phi}_s$) and quarton ($\tilde{\phi}_{q\Sigma}$) are biased (see Appendix \ref{section:flux-calibration} for the calibration procedure).

Note that when the transmons are strongly cross-Kerr coupled, i.e. $\chi \hat{a}^\dagger\hat{a}\hat{b}^\dagger\hat{b}$ with $|\chi| \gg 0$, the device exhibits an unusual phenomenon where both resonators can be used to readout either transmon. This is because the capacitive coupling $g$ of transmon A(B) to its $\Delta$ frequency-detuned resonator A(B) hybridizes their modes, this can be approximated as $\hat{a}(\hat{b}) \rightarrow \hat{a}(\hat{b}) + \frac{g}{\Delta}\hat{a}_{ro}(\hat{b}_{ro})$ where $\hat{a}_{ro} (\hat{b}_{ro})$ are annihilation operators of readout resonator A(B). The hybridization imparts the usual dispersive shifts, $\chi_{d,a} \hat{a}^\dagger \hat{a} \hat{a}_{ro}^\dagger \hat{a}_{ro}$ and $\chi_{d,b} \hat{b}^\dagger \hat{b} \hat{b}_{ro}^\dagger \hat{b}_{ro}$, but also an additional non-dispersive cross-Kerr $\chi_n$ with approximately:
\begin{equation}
\begin{aligned}
& \chi \hat{a}^\dagger\hat{a}\hat{b}^\dagger\hat{b} \\
\rightarrow & \chi (\hat{a}^\dagger + \frac{g}{\Delta}\hat{a}_{ro}^\dagger)(\hat{a} + \frac{g}{\Delta}\hat{a}_{ro})(\hat{b}^\dagger + \frac{g}{\Delta}\hat{b}_{ro}^\dagger)(\hat{b} + \frac{g}{\Delta}\hat{b}_{ro}) \\
= & \chi \hat{a}^\dagger\hat{a}\hat{b}^\dagger\hat{b} + \chi_n \hat{a}^\dagger\hat{a}\hat{b}_{ro}^\dagger\hat{b}_{ro} + \chi_n \hat{b}^\dagger\hat{b}\hat{a}_{ro}^\dagger\hat{a}_{ro} + \dots
\end{aligned}
\label{eq: non-disp-ro}
\end{equation}
Unlike the usual dispersive shift which for a transmon is proportional to its self-Kerr \cite{KochTransmon} ($\chi_{d,a(b)} \propto K_{a(b)}$) and thus vanishes to first-order when the transmon is linearized ($K \approx 0$), the non-dispersive $\chi_n$ is independent of transmon self-Kerrs and can thus be leveraged to readout linearized transmons. 
\subsection{Spectroscopy}
\begin{figure*}
    \includegraphics[width=1\textwidth]{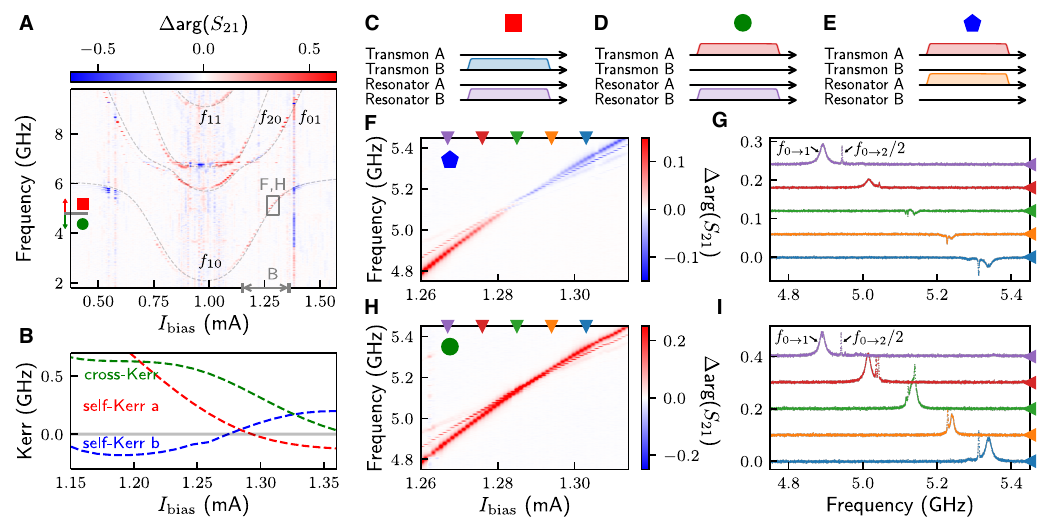}
    \caption{\textbf{Transmon self-Kerr tuning via quarton coupler flux bias.}  (\textbf{A})~Two-tone spectroscopy of the device with theory fit (grey dashed) overlaid. (\textbf{B})~Self- and cross-Kerr of transmons A and B at different quarton flux bias, extracted from the theory fit. Transmon A reaches zero self-Kerr at approximately $I_{\text{bias}}=1.285$ mA. (\textbf{C-E})~Pulse sequences for two-tone spectroscopy, labeled by colored shapes.  (\textbf{F-I})~High power two-tone spectroscopy near zero self-Kerr ($I_{\text{bias}}=1.285$ mA) with pulse sequences E (for F-G) and D (for H-I). Clear signature of linearization can be observed, with peaks converging in both spectroscopies and the dispersive shift changing signs in panel F. Panels G, I display respective line-cuts of F, H (at $I_{\text{bias}}$ labeled by colored triangles), where single photon $f_{0\rightarrow1}$ and multi-photon $f_{0\rightarrow2}/2$ transitions are visible. The phase of successive line-cuts are plotted with a constant offset for visual clarity.}
    \label{fig: exp_2}
\end{figure*}

We obtain the circuit's eigenenergy spectrum as a function of quarton flux bias (Fig.~\ref{fig: exp_2}A) by performing standard two-tone spectroscopy \cite{gao2021practical} while sweeping $I_\text{bias}$ (a proxy for quarton flux bias $\tilde{\phi}_{q\Sigma}$, see Appendix \ref{section:flux-calibration} for details). Since transmon B is designed to have a higher frequency, 
we apply the drive through transmon B's drive line (Fig.~\ref{fig: exp_2}C) when performing spectroscopy at high frequency.
For lower frequency spectroscopy, we instead drive transmon A, which is designed with a lower frequency (Fig.~\ref{fig: exp_2}D). In both cases we use resonator B for readout because resonator A (at 6.837 GHz) is accidentally near-resonant with transitions at certain $I_\text{bias}$.
Fig.~\ref{fig: exp_2}A reveals several transition frequencies (labeled $f_{n_A n_B}$ on the plot by the excitation number in transmon A(B) denoted $n_{A(B)}$) of our system. By numerically solving for the eigenenergies of the circuit and fitting the Josephson energies of each JJ as free parameters (see Appendix \ref{sec:exp_fit} for details), we obtain good agreement with the spectroscopy results (gray dashed lines). 

From the theory fit, we compute expected self- and cross-Kerrs as shown in Fig.~\ref{fig: exp_2}B. Around the bias point $I_{\text{bias}}=1.285$ mA, the model predicts the desired nonlinear light-matter coupling properties with near-zero self-Kerr for transmon A, non-zero self-Kerr for transmon B, and a large cross-Kerr between them. We also identify bias points such as $I_{\text{bias}}=1.224$ mA where both transmons behave like large self-Kerr qubits, whose extremely large cross-Kerr coupling is ideal for matter-matter nonlinear coupling (also known as $ZZ$ \cite{residualZZ} or Ising \cite{IsingJRM} or longitudinal interaction \cite{trimon2017}: $\frac{\chi}{4}\hat{\sigma}_{z,a}\hat{\sigma}_{z,b}$).

In Fig.~\ref{fig: exp_2}F-I, we zoom in and more closely examine the flux bias near $I_{\text{bias}}=1.285$ mA where transmon A is linearized. We perform standard high-power two-tone spectroscopy so the multi-photon transitions that reveal transmon anharmonicity can be excited \cite{gao2021practical}. In Fig.~\ref{fig: exp_2}F, we drive transmon A and resonantly probe the dispersively-coupled resonator A (see Fig.~\ref{fig: exp_2}E). We observe a clear sign change in readout phase indicating a corresponding sign change in the underying dispersive shift between resonator A and transmon A. This is expected as a resonator's dispersive shift with a transmon (when $\Delta \gg K$) is directly proportional to the transmon's self-Kerr \cite{KochTransmon} ($\chi_{d} \propto K$), and we also observe a concurrent change in self-Kerr anharmonicity, most clearly-observed in Fig.~\ref{fig: exp_2}G where we plot the line-cuts of Fig.~\ref{fig: exp_2}F with constant phase offsets. Here, we see higher-order transition peaks (most visibly, $f_{0\rightarrow2}/2$) move from above to below the $f_{0\rightarrow1}$ peak and converge in the middle, near the theory-predicted zero-Kerr point $I_{\text{bias}}=1.285$ mA.
At this point, the transmon A peak is almost invisible to its dispersively-coupled readout resonator A, consistent with the prediction that the dispersive shift goes to zero at linearization.
\begin{table*}
\begin{center}
\caption{\label{tab:exp_table_summary}Summary of frequencies (MHz) and coherence times ($\mu$s) of both transmons at operating point $I_{\text{bias}}=1.285$ mA where transmon A has near-zero anharmonicity.}
\addtolength{\tabcolsep}{+5pt}
\begin{tabular}{|c|c|c|c|c|c|c|}
\hline & $\chi/2\pi$ (MHz) & $f_{0\rightarrow1}$ (MHz) & $f_{1\rightarrow2} - f_{0\rightarrow1}$ (MHz) & $f_{2\rightarrow3} - f_{0\rightarrow1}$ (MHz) & $T_1$ ($\mu$s)  & $T_2^E$ ($\mu$s)\\
\hline \hline Transmon A & 366.25 $\pm$ 0.84 & 5105.02 $\pm$ 0.04 & 0.76 $\pm$ 0.08 & -60.46 $\pm$ 0.41 & 10.61 $\pm$ 0.27 & - \\
\hline Transmon B & 365.69 $\pm$ 0.36 & 7542.42 $\pm$ 0.02 & 25.44 $\pm$ 0.11 & 259.10 $\pm$ 0.35 & 8.55 $\pm$ 0.14 & 2.23 $\pm$ 0.04 \\
\hline 
\end{tabular}
\addtolength{\tabcolsep}{-5pt}
\end{center}
\end{table*}
\begin{figure*}
    \includegraphics[width=1\textwidth]{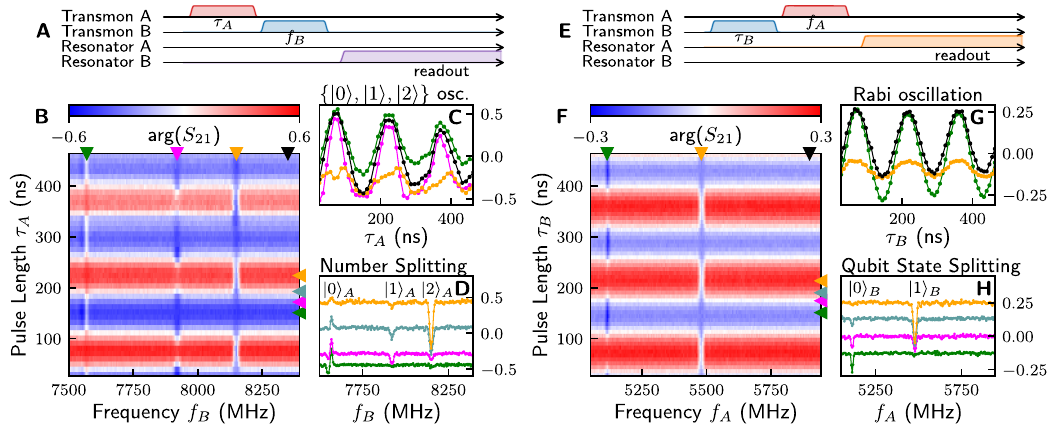}
    \caption{\textbf{Near-ultrastrong nonlinear coupling between linearized transmon A (light) and transmon qubit B (matter).} (\textbf{A}) Pulse diagram:  resonant pulse of length $\tau_A$ driving linearized transmon A followed by pulse of frequency $f_B$ driving transmon B and readout with resonator B. (\textbf{B}) Readout resonator B response as a function of $\tau_A$ and $f_B$. (\textbf{C}) Vertical line-cuts of panel B showing Rabi-like oscillation. (\textbf{D}) Horizontal line-cuts of panel B showing photon-number splitting of transmon B transition by transmon A's excitation number $\{\ket{0},\ket{1},\ket{2}\}_A$. (\textbf{E}) Pulse diagram:  resonant pulse of length $\tau_B$ driving transmon qubit B followed by pulse of frequency $f_A$ driving linearized transmon A and readout with resonator A. (\textbf{F}) Readout resonator A response as a function of $\tau_B$ and $f_A$. (\textbf{G}) Vertical line-cuts of panel F showing Rabi oscillation. (\textbf{H}) Horizontal line-cuts of panel F showing splitting of transmon A transition by transmon B's qubit states $\{\ket{0},\ket{1}\}_B$.}
    \label{fig: exp_3}
\end{figure*}

We verify that the disappearance of transmon A (in Fig.~\ref{fig: exp_2}F) is due to its linearization by repeating high-power two-tone spectroscopy with resonator B instead (see Fig.~\ref{fig: exp_2}D). As derived previously (see Eq.~(\ref{eq: non-disp-ro})), there exists a non-dispersive cross-Kerr $\chi_n$ between transmon A and resonator B which does not depend on transmon A's anharmonicity $K_a$. As predicted, the resulting spectroscopy (Fig.~\ref{fig: exp_2}H-I) shows the same convergence of higher-order transitions at the linearization point but has a strong transmon A signal even when it is linearized. In fact, among the Fig.~\ref{fig: exp_2}I line-cuts, the transmon A peak is the strongest at the linearization point (green) because more energy levels can be excited (higher $\langle\hat{a}^\dagger \hat{a}\rangle$) for an overall larger readout shift ($\chi_n\langle\hat{a}^\dagger \hat{a}\rangle$) on resonator B. We also see that the phase shifts in Fig.~\ref{fig: exp_2}H-I are all positive, in agreement with the prediction of Eq.~(\ref{eq: non-disp-ro}) that $\chi_n \propto \chi$ and the quartonic $\chi$ between transmon A and B is positive (see Fig.~\ref{fig: exp_2}B). We note that this new non-local, non-dispersive cross-Kerr interaction between a transmon and a spatially-separated and geometrically-uncoupled resonator may have further applications in novel readout or remote-entanglement schemes.

\subsection{Near-ultrastrong light-matter nonlinear coupling}
We now demonstrate near-ultrastrong nonlinear coupling between transmon B and the linearized transmon A by operating at the linearization point ($I_{\text{bias}}=1.285$ mA) found previously. Table \ref{tab:exp_table_summary} shows the transition frequencies and coherence times of both transmons at this operating point (see Appendix \ref{sec:transmon_freq_T1T2} for details). 
We note that transmon A has a near-zero measured self-Kerr anharmonicity of $0.76\pm0.08$ MHz, on par with or lower than other experimental self-Kerr anharmonicities of light-like resonator modes reported in literature \cite{Holland2015, Dassonneville2020}, which allows its non-qubit ($\ket{i}$, $i>1$) states to be excited under resonant drive pulses. Transmon A's linearization is limited by its higher-order six-wave-mixing ($\hat{a}^{\dagger3} \hat{a}^3$) anharmonicity of $-60.46 \pm 0.41$ MHz, so for resonant drives with low amplitudes (Rabi frequency $\Omega_R/2\pi \ll 60$ MHz), this six-wave-mixing anharmonicity suppresses excitation beyond the first 3 levels ($\{\ket{0},\ket{1},\ket{2}\}$). 
\begin{figure*}
    \includegraphics[width=1\textwidth]{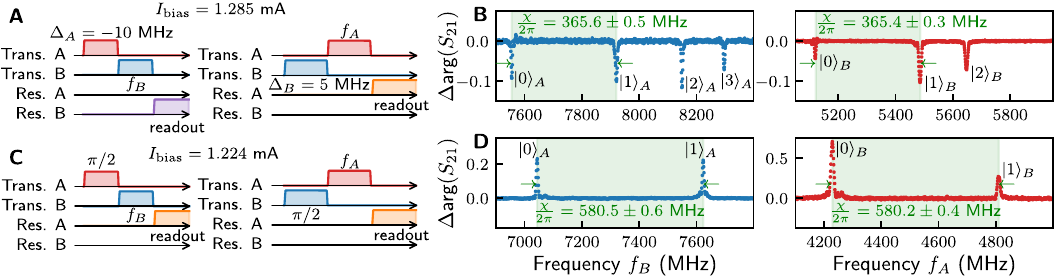}
    \caption{\textbf{Matter-matter and simulated light-light nonlinear coupling.} (\textbf{A}) Pulse diagrams for simulated light-light nonlinear coupling experiments at $I_{\text{bias}}=1.285$ mA. The initial transmon A(B) drive pulse is frequency detuned from ($f_{0\rightarrow1}$) resonance by $\Delta_{A(B)} = -10 (+5) $ MHz to better excite higher energy level transitions. (\textbf{B}) Spectroscopies showing photon-number splitting of both transmons' transition, a key signature of cross-Kerr between two photon modes. First 4 levels of transmon A and 3 levels of transmon B are visible with $\{\ket{0},\ket{1}\}$ splitting of $\chi/2\pi = \text{365.6(4)} \pm \text{0.5(3)}$ MHz for transmon A(B). Left, right panels of number splitting results are obtained from respective left, right pulse diagrams (panel A). (\textbf{C}) Pulse diagrams for matter-matter nonlinear coupling experiments at $I_{\text{bias}}=1.224$ mA. The initial transmon A(B) drive is a resonant $\pi/2$ pulse. (\textbf{D}) Spectroscopies showing qubit state splitting of both transmons' transition, as expected for cross-Kerr between two qubit modes. Measured $\{\ket{0},\ket{1}\}$ splittings of $\chi/2\pi = \text{580.5(2)} \pm \text{0.6(4)}$ MHz for transmon A(B). Left, right panels of qubit state splitting results are obtained from respective left, right pulse diagrams (panel C).  }
    \label{fig: exp_4}
\end{figure*}

To confirm this, we apply the pulse sequence shown in Fig.~\ref{fig: exp_3}A:
we first resonantly drive the linearized transmon A with a low $\Omega_R$ pulse of varying duration $\tau_A$, 
then apply a pulse of varying frequency $f_B$ to transmon B, and finally end by probing resonator B.
We plot in Fig.~\ref{fig: exp_3}B the resulting phase of resonator B as a function of the two swept variables $\tau_A$ and $f_B$. We note again that resonator B is dispersively-coupled with $\chi_d$ to transmon B and non-dispersively cross-Kerr coupled with $\chi_n$ to transmon A, so resonator B's phase conveniently encodes both transmons' population.
Examining Fig.~\ref{fig: exp_3}B and its line-cuts in  Fig.~\ref{fig: exp_3}C-D, we observe Rabi-like oscillation along time $\tau_A$ and varying splitting of transmon B transition that indicates the Rabi-like oscillation is between states $\{\ket{0},\ket{1},\ket{2}\}$ of transmon A as expected. 
We emphasize that the drive-dependent photon-number splitting spectrum in Fig.~\ref{fig: exp_3}D is a defining signature of strong light-matter nonlinear coupling \cite{number_splitting}. Here the $\ket{0}_A$ and $\ket{1}_A$ transitions are split by a cross-Kerr $\chi/2\pi = 366.25 \pm 0.84$ MHz (see Appendix \ref{sec:LM-chi-extraction} for details), which is more than four times larger than the state of the art \cite{Inomata2012}. 
The higher photon-number $\ket{2}_A$ transition exhibits lower cross-Kerr, which results from parasitic $\hat{a}^\dagger\hat{a} (\hat{a}^\dagger \hat{b} + \hat{b}^\dagger \hat{a})$ interactions originating from $\phi_a^3 \phi_b$ terms in the quarton coupling potential $(\phi_a-\phi_b)^4$. These interactions increase the effective linear coupling rate for higher photon-number states, thereby lowering cross-Kerr in agreement with theoretical predictions (see Appendix \ref{sec:exp_fit} for details). 

As a complementary experiment, we probe the system response when the nonlinear transmon B is excited first, followed by spectroscopy on linear transmon A and readout through resonator A, as described in Fig.~\ref{fig: exp_3}E. Similar to before, we plot the phase of resonator A in Fig.~\ref{fig: exp_3}F, and show corresponding vertical and horizontal line-cuts in Fig.~\ref{fig: exp_3}G and Fig.~\ref{fig: exp_3}H, respectively. Since transmon B has much larger self-Kerr anharmonicity of $25.44 \pm 0.11$ MHz (Table \ref{tab:exp_table_summary}) compared to the drive amplitude, we see a Rabi oscillation expected of driven qubits in Fig.~\ref{fig: exp_3}G. We also observe in Fig.~\ref{fig: exp_3}H a splitting of transmon A's transition by transmon B's qubit states $\{\ket{0},\ket{1}\}_B$, with the relative strength of each peak varying in accordance with expected qubit population oscillation during the Rabi cycle.
We again extract the cross-Kerr from the $\{\ket{0},\ket{1}\}_B$ splitting to be $\chi=365.69\pm0.36$ MHz (see Appendix \ref{sec:LM-chi-extraction} for details). The two cross-Kerr values show excellent agreement within measurement uncertainty and average to $\chi=366.0\pm0.5$ MHz, leading to $\tilde{\eta}=(4.852\pm0.006)\times10^{-2}$ in the near-ultrastrong nonlinear light-matter coupling regime.

\subsection{Simulated light-light nonlinear coupling}
Transmon B exhibits qubit-like behavior under a weak, resonant drive in Fig.~\ref{fig: exp_3}, but its small self-Kerr anharmonicity can be exploited under a strong, off-resonant drive to excite higher levels and exhibit resonator-like behavior instead. 
As shown in Fig.~\ref{fig: exp_4}A, we repeat the experiment in Fig.~\ref{fig: exp_3} but now apply the first pulse with larger amplitude and a detuning of $\Delta_{A(B)}=-10 (+5)$ MHz from transmon A(B)'s $f_{0\rightarrow1}$ transition (see Appendix \ref{sec:light-light-in-time} for detailed time domain results). 
This allows us to simulate the regime where both transmons are linearized or light-light nonlinear coupling. 
The resulting spectroscopy in Fig.~\ref{fig: exp_4}B shows clear signature of photon-photon cross-Kerr \cite{Holland2015}, with number splitting for both transmons, by $\{\ket{0},\ket{1},\ket{2},\ket{3}\}_A$ and $\{\ket{0},\ket{1},\ket{2}\}_B$, respectively. As expected for the same device operating point, the extracted $\chi$ is the same as in Table \ref{tab:exp_table_summary}. Compared to state-of-the-art $\chi/2\pi = 2.59$ MHz 
\cite{Holland2015}, our simulated light-light coupling demonstrates more than two orders of magnitude increase in 
$\chi$.
We emphasize that with a greater range of flux-tunability or more precise parameter targeting in fabrication, our quartonic architecture is capable \cite{quartonPRL} of demonstrating light-light nonlinear coupling with both transmons linearized to state-of-the-art levels ($\leq 4 $ MHz \cite{Holland2015}).

\subsection{Matter-matter nonlinear coupling}
To explore the regime of maximal  nonlinear coupling with our device, we follow theory predictions of Fig.~\ref{fig: exp_2}B and flux-bias the gradiometric quarton coupler to $I_{\text{bias}}=1.224$ mA. This coincides with a matter-matter coupling regime where both transmons have high self-Kerr anharmonicity and thus behave as qubits or artificial atoms (see Appendix \ref{sec:transmon_freq_T1T2} for detailed qubit properties). We then measure cross-Kerr coupling by performing the experiment outlined in Fig.~\ref{fig: exp_4}C: applying first a $\pi/2$ pulse to one qubit, followed by spectroscopy of the other qubit and readout. The spectroscopy results in Fig.~\ref{fig: exp_4}D shows the expected qubit state splitting, with an extremely large extracted cross-Kerr of $\chi/2\pi = \text{580.5(2)} \pm \text{0.6(4)}$ for transmon A(B). The averaged $\chi/2\pi = 580.3 \pm 0.4 $ MHz is, to the best of our knowledge, the largest $ZZ$ coupling rate between two coherent qubits of any physical platform, and is equivalent to a $CZ$ gate time of 0.86 ns. Here we exclude comparison with annealer architectures such as \cite{dwave} that lack measurable qubit coherence. 

\section{Conclusion}
We experimentally demonstrate a quartonic approach to nonlinear coupling, capable of both large cross-Kerr coupling and self-Kerr cancellation which can linearize transmon qubits into nearly-linear resonator modes. This allows us to show the first near-ultrastrong nonlinear light-matter coupling with $\tilde{\eta} = (4.852\pm0.006)\times10^{-2}$ and $\chi/2\pi = 366.0 \pm 0.5 $ MHz. 
We also show that this large $\chi$ persists in a simulated regime of light-light nonlinear coupling \cite{Holland2015}. At another operating point, we measure a matter-matter nonlinear coupling of $\chi/2\pi = 580.3 \pm 0.4 $ MHz, which is (to the best of our knowledge) the largest cross-Kerr ($ZZ$) coupling between two coherent qubits.
Our work motivates future explorations in subsequent regimes of ultrastrong and deep-strong nonlinear light-matter coupling, and 
may lead to orders of magnitude improvements in fundamental quantum-information operations such as qubit gates and readout.

\vspace{2em}

\section{Acknowledgment}
The authors thank Jens Koch, Alessandro Miano, Leon Ding, Max Hays, David Rower, Tianpu Zhao, Terry Orlando, William Oliver, Mahdi Naghiloo, and David Toyli for fruitful discussions and insightful comments.

This research was supported in part by the Army Research Office under Award No. W911NF-23-1-0045, the AWS Center for Quantum Computing, and the MIT Center for Quantum Engineering via support from the Laboratory for Physical Sciences under Contract No. H98230-19-C-0292. This material is based upon work supported under Air Force Contract No. FA8702-15-D-0001. Any opinions, findings, conclusions, or recommendations expressed in this material are those of the author(s) and do not necessarily reflect the views of the U.S. Air Force. Y.Y. acknowledges support from the IBM PhD Fellowship and the NSERC Postgraduate Scholarship. J.B.K acknowledges support from the Alan L. McWhorter (1955) Fellowship. A.Y. acknowledges support from the NSF Graduate Research Fellowship. G.C. acknowledges support from the Harvard Graduate School of Arts and Sciences Prize Fellowship.

\textbf{Author Contributions:} Y.Y. conceived the idea, designed the chip, performed the experiment and simulations. J.B.K. assisted in experiment and performed simulations.  A.Y., G.C., M.T. and A.Z. assisted in experimental setup. M.G., B.M.N., and H.S. fabricated the device. K.P.O., K.S., and M.S. supervised and guided the project. Y.Y., J.B.K, A.Y. wrote the manuscript with input from all authors. All authors discussed the results and provided feedback to the manuscript. 

\textbf{Competing interests:} The authors declare that they have no competing interests.

\textbf{Data and materials availability:} All data are available in the manuscript or the supplementary materials.

\appendix
\section{\label{sec:methods}Device and setup}
The device was fabricated with thin-film aluminum on silicon substrate. Superconducting air-bridges were included along coplanar waveguide sections to suppress slot-line modes.

\begin{figure}    
\includegraphics[width=\columnwidth]{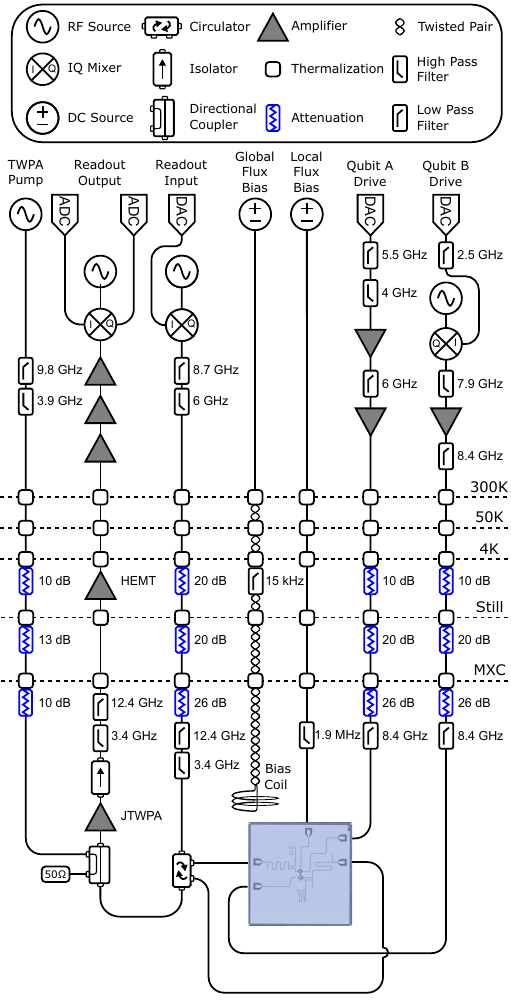}
    \caption{\textbf{Diagram of experimental setup.} The cryogenic setup includes local and global DC flux-bias lines, two qubit drive lines, and one shared readout line. Component symbol meanings are enumerated at the top and relevant values are shown next to the components in the diagram.}
    \label{fig:fridge}
\end{figure}

The experimental diagram is shown in Fig.~\ref{fig:fridge}. The experiment was conducted in a Bluefors LD400 dilution
refrigerator with a base temperature of approximately 20 mK at the mixing
chamber (MXC). The packaged device was enclosed at the MXC by two nested shields: an inner superconducting
aluminum shield, and an outer Cryoperm shield. Continuous-wave probe tones were applied with a Rohde and Schwarz SGS100A signal generator. Microwave readout and drive pulses were
applied by a ZCU111 RFSoC FPGA with QICK \cite{QICK} programming. Global and local flux bias were applied through Keithley 6220 DC current sources.

\section{Flux analysis and calibration}\label{section:flux-calibration}
\begin{figure}
    \includegraphics[width=0.5\textwidth]{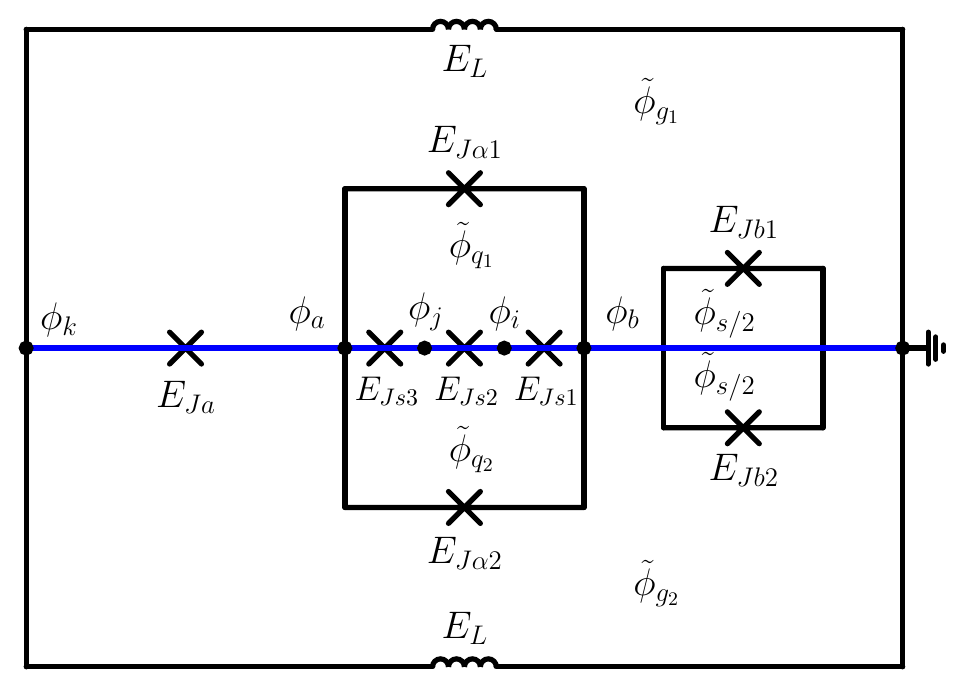}
    \caption{\textbf{Full circuit diagram including all loops.} Loops through ground, neglected in main text Fig.~\ref{fig: exp_1}C, are included. The spanning tree chosen is shown in blue. Relevant nodal (represented as dot) phase $i$ is labeled by $\phi_i$ (next to each respective dot) and flux bias $j$ is labeled by $\tilde{\phi}_j$ (in each respective loop).}
    \label{fig:loops}
\end{figure}
To understand the effect of flux bias on all the loops of the circuit, we analyze the circuit diagram shown in Fig.~\ref{fig:loops}, which includes the two loops through ground by treating the ground path as an inductor with energy $E_L$.

Drawing a spanning tree (blue) through the center of the circuit, we can assign flux biases and write the potential as
\begin{equation}
\begin{aligned}
& U=-E_{J b_1} \cos \left(\phi_b-\frac{\tilde{\phi}_s}{2}\right)-E_{J b_2} \cos \left(\phi_b+\frac{\tilde{\phi}_s}{2}\right) \\
& -E_{J s_1} \cos \left(\phi_b-\phi_i\right)-E_{J s_2} \cos \left(\phi_i-\phi_j\right)-E_{Js_3} \cos \left(\phi_j-\phi_a\right) \\
& -E_{J \alpha_1} \cos \left(\phi_a-\phi_b-\tilde{\phi}_{q_1}\right)-E_{J \alpha_2} \cos \left(\phi_a-\phi_b+\tilde{\phi}_{q_2}\right) \\
& -E_{Ja} \cos \left(\phi_a-\phi_k\right)+\frac{1}{2} E_L\left[\left(\phi_k-\tilde{\phi}_{g_1}\right)^2+\left(\phi_k+\tilde{\phi}_{g_2}\right)^2\right]
\end{aligned}
\end{equation}

Assuming the ground path has a very large $E_L$ (low inductance), then $\phi_k$ will be localized at the point which minimizes the $E_L$ term. Rewriting this term as
\begin{equation}
\begin{aligned}
    & \frac{1}{4} E_L\left[\left(2 \phi_k+\tilde{\phi}_{g_2}-\tilde{\phi}_{g_1}\right)^2+\left(\tilde{\phi}_{g_1}+\tilde{\phi}_{g_2}\right)^2\right] \\
    & =\frac{1}{4} E_L\left[\left(2 \phi_k-\tilde{\phi}_{g_\Delta}\right)^2+\phi_{g_\Sigma}^2\right],
\end{aligned}
\end{equation}
 where $\tilde{\phi}_{g_\Delta} = \tilde{\phi}_{g_2} - \tilde{\phi}_{g_1}$ and $\tilde{\phi}_{g_\Sigma} = \tilde{\phi}_{g_2} + \tilde{\phi}_{g_1}$, we see that the energy minimum is at $\phi_k = \tilde{\phi}_{g_\Delta}/2$. Plugging this in, dropping constant terms, and rearranging leads to our final expression for the potential energy, 
\begin{equation}\label{eq:full-potential}
\begin{aligned}
& U=-(E_{J b_1} +E_{J b_2})\cos\left(\tilde{\phi}_s/2\right)\cos \left(\phi_b\right) \\
& - (E_{J b_2}-E_{J b_1})\sin\left(\tilde{\phi}_s/2\right) \sin \left(\phi_b\right) \\
& -E_{J s_1} \cos \left(\phi_b-\phi_i\right)\\
& -E_{J s_2} \cos \left(\phi_i-\phi_j\right) \\
& -E_{Js_3} \cos \left(\phi_j-\phi_a\right) \\
& -(E_{J \alpha_1} + E_{J\alpha_2})\cos(\frac{\tilde{\phi}_{q_\Sigma}}{2}) \cos \left(\phi_a-\phi_b-\frac{\tilde{\phi}_{q_\Delta}}{2}\right) \\
& -(E_{J \alpha_2} - E_{J\alpha_1})\sin(\frac{\tilde{\phi}_{q_\Sigma}}{2}) \sin \left(\phi_a-\phi_b-\frac{\tilde{\phi}_{q_\Delta}}{2}\right) \\
& -E_{Ja} \cos \left(\phi_a-\tilde{\phi}_{g_\Delta}/2\right)
\end{aligned}
\end{equation}
where $\tilde{\phi}_{q_\Sigma} = \tilde{\phi}_{q_1} + \tilde{\phi}_{q_2}$ and $\tilde{\phi}_{q_\Delta} = \tilde{\phi}_{q_1} - \tilde{\phi}_{q_2}$. This means our system depends on four external fluxes:  $\tilde{\phi}_{s}$,$  \tilde{\phi}_{g_\Delta}$,$\tilde{\phi}_{q_\Sigma}$, and $ \tilde{\phi}_{q_\Delta}$.

\begin{figure*}
    \includegraphics[width=1.0\textwidth]{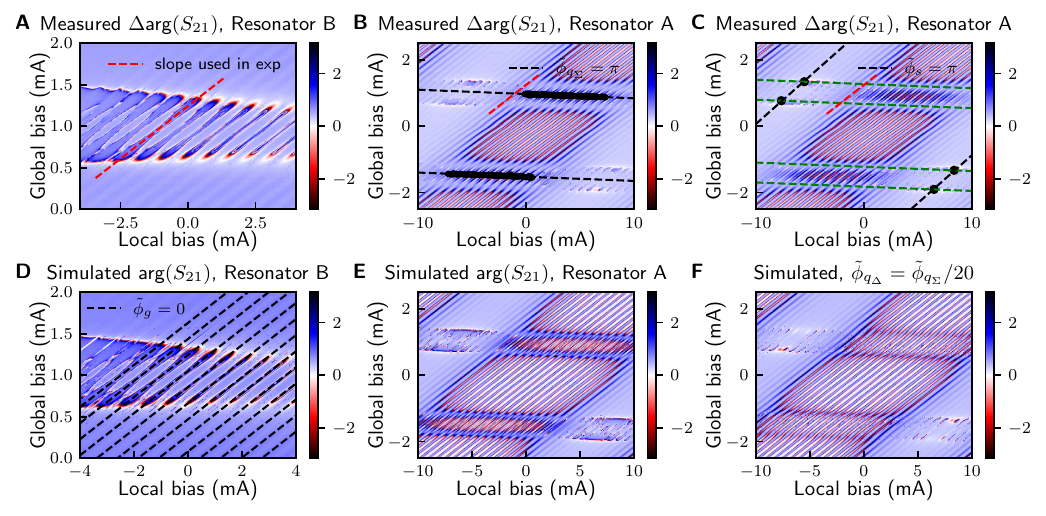}
    \caption{\textbf{Flux calibration procedure. (A)} The red line is chosen by aligning to the narrow diagonal features which should be approximately where $\tilde{\phi}_g = 0$. \textbf{(B)} Quarton flux $\tilde{\phi}_{q_\Sigma}$ calibration. Black dots indicate points where the resonator response was symmetric along a point of constant $\tilde{\phi}_g$, and black dashed lines are fits to these points, indicating $\tilde{\phi}_{q_\Sigma}$ is approximately $\pi$. \textbf{(C)} SQUID flux $\tilde{\phi}_{s}$ calibration. Blue dots indicate points where $\tilde{\phi}_s = \pi$, found by fitting a Gaussian to $|S_{21}|$ along the constant $\tilde{\phi}_{q_\Sigma}$ lines indicated in green. Black lines, fit using an average of the blue points, indicate where $\tilde{\phi}_s = \pi$. \textbf{(D)} Simulated resonator B response. Black dashed lines indicate where $\tilde{\phi}_g = 0$ in simulation. \textbf{(E)} Simulated resonator A response with $\tilde{\phi}_{q_\Delta} = 0$. \textbf{(F)} Simulated resonator response with $\tilde{\phi}_{q_\Delta} = \tilde{\phi}_{q_\Sigma}/20$.}
    \label{fig: flux-calibration}
\end{figure*}

We measure the periodicities of these fluxes by probing resonator A at a constant frequency near resonance and sweeping both the local and global flux biases. From the resulting $S_{21}$ maps (Fig.~\ref{fig: flux-calibration}), we can identify periodic features belonging to each flux bias. To find a line where $\tilde{\phi}_{g_\Delta}=0$, we assume that $E_{Ja_1} = E_{Ja_2}$ and $\tilde{\phi}_{q_\Delta} = 0$ (both expected from design symmetry). This means that taking $\tilde{\phi}_{g_\Delta} 
 \rightarrow -\tilde{\phi}_{g_\Delta}$ has the effect of mirroring potential about each coordinate, so the eigenenergies of the system should be symmetric in $\tilde{\phi}_{g_\Delta}$ about $\tilde{\phi}_{g_\Delta}=0$. Since $\tilde{\phi}_{g_\Delta}$ varies with applied current much faster than any other flux in our system, we align to the sharp features in Fig.~\ref{fig: flux-calibration}A to experimentally find the line with $\tilde{\phi}_{g_\Delta}=0$. This red line in Fig.~\ref{fig: flux-calibration}A has equation in terms of global and local bias ($I_g$, $I_l$):
 \begin{equation}
     I_g = 0.254 \times I_l + 1.243.
\label{eq: diagonal}
 \end{equation}
Since changing $I_g$ has approximately the effect of changing only the quarton flux bias $\tilde{\phi}_{q_\Sigma}$, we define $I_\text{bias}$ as the $I_g$ of points ($I_g$, $I_l$) along the Eq.~(\ref{eq: diagonal}) line.


 Sweeping the bias currents further (Fig.~\ref{fig: flux-calibration}B), we see two broad horizontal stripes corresponding to $\tilde{\phi}_{q_\Sigma}$. We find their period in both directions by assuming that along lines of constant $\tilde{\phi}_{g_\Delta}$, there will be symmetry around the point where $\tilde{\phi}_{q_\Sigma} = \pi$. For many lines of constant $\tilde{\phi}_{g_\Delta}$, we estimate these $\tilde{\phi}_{q_\Sigma} = \pi$ points by reflecting the complex $S_{21}$ about each point along the constant $\tilde{\phi}_{g_\Delta}$ line and computing the dot product between the orignal and reflected lines. This yields a series of points approximating the two lines of $\tilde{\phi}_{s}$, and we then use a Huber regression to fit these lines, which determines the period and phase of $\tilde{\phi}_{q_\Sigma}$. Because this period is already on the order of the range of currents we used, and because the quarton loops are approximately symmetric, we assume that $\tilde{\phi}_{q_\Delta} = 0$ everywhere. Finally, we determine the periodicity of $\tilde{\phi}_{s}$ by noting that along lines of constant $\tilde{\phi}_{q_\Sigma}$, and averaging over the fast oscillation caused by $\tilde{\phi}_{g_\Delta}$, we should have symmetry about the $\tilde{\phi}_{s} = \pi$ points. The oscillations of $\tilde{\phi}_{g_\Delta}$ prevent us from directly computing a dot product as we did for $\tilde{\phi}_{q_\Sigma}$, so instead we identify four lines of constant $\tilde{\phi}_{q_\Sigma}$ which each pass through a visible feature centered at $\tilde{\phi}_{s} = \pi$. We fit a Gaussian to these features (taking the magnitude, $|S21|$) and use the resulting four points (black in Fig.~\ref{fig: flux-calibration}C) to compute the periodicity and phase of $\tilde{\phi}_{s}$.

We verify these estimates by numerically simulating the resonator $\arg(S_{21})$ as a function of local and global bias, assuming our estimated periodicities. To expedite this time-intensive simulation, we neglect junction asymmetry as well as the internal modes of the series junctions (see the potential in Eq.~(\ref{eq:potential})). This introduces a small error in the simulated eigeneneries which does not affect the flux periodicities we are primarily interested in. We use the junction energies found in Table~\ref{tab:JJ_theory}, where for any junctions designed to be identical, we assume the mean value for both of them. We include the resonator as a mode in the system, numerically diagonalizing the Hamiltonian to find the resonator frequency at each point. From the resonator frequency, we can compute \cite{mcrae2020}
\begin{equation}
    S_{21}(f) = 1 - \frac{2\frac{Q}{Q_c}}{1+2iQ\frac{f-f_0}{f_0}}
\end{equation}
where $f_0$ is the resonator frequency, $Q_c$ is the coupling quality factor of the resonator to its transmission line, and $Q=(\frac{1}{Q_c} + \frac{1}{Q_i})^{-1}$ where $Q_i$ is the internal quality factor of the resonator. 
In Fig.~\ref{fig: flux-calibration}D, we simulate the region used in Fig.~\ref{fig: flux-calibration}A, noting that indeed the $\tilde{\phi}_{g_\Delta} = 0$ lines can be associated with symmetries in the data. In Fig.~\ref{fig: flux-calibration}E, we simulate the region of Fig.~\ref{fig: flux-calibration}B,C and see good qualitative agreement. Finally, to verify the validity of assuming $\tilde{\phi}_{q_\Delta} = 0$, we perform the same simulation by assuming the $\tilde{\phi}_{q_\Delta}$ period is 20 times larger than the $\tilde{\phi}_{q_\Sigma}$ period, in both the global and local bias currents. The resulting pattern contains prominent differences near the  lines of $\tilde{\phi}_{q_\Sigma} = \pi$ which do not appear in the experimental flux sweep, indicating that the experimental period of $\tilde{\phi}_{q_\Delta}$ must be even larger than 20 times the $\tilde{\phi}_{q_\Sigma}$ period.

\section{\label{sec:exp_fit}Device parameter fit}
\begin{table*}
\begin{center}
\caption{\label{tab:freqs_vs_theory} Comparison of measured (from Fig.~\ref{fig: exp_4} main text) vs. theory transition frequencies.}
\addtolength{\tabcolsep}{+5pt}
\begin{tabular}{|c|c|c|c|c|}
\hline Transition & $I_{\text{bias}}$ (mA) & Measured (MHz) & Theory (MHz) & Error (\%) \\
\hline 
\hline $f_{10} - f_{00}$ & 1.285 & 5121.27 $\pm$ 0.25 & 5064.72 & 1.10\\
\hline $f_{11} - f_{01}$ & 1.285 & 5486.70 $\pm$ 0.16 & 5461.57 & 0.46\\
\hline $f_{12} - f_{02}$ & 1.285 & 5649.56 $\pm$ 0.25 & 5640.7 & 0.16\\
\hline $f_{01} - f_{00}$ & 1.285 & 7554.41 $\pm$ 0.38 & 7512.94 & 0.55\\
\hline $f_{11} - f_{10}$ & 1.285 & 7920.04 $\pm$ 0.32 & 7909.80 & 0.13\\
\hline $f_{21} - f_{20}$ & 1.285 & 8148.99 $\pm$ 0.27 & 8150.86 & 0.02\\
\hline $f_{31} - f_{30}$ & 1.285 & 8296.27 $\pm$ 0.40 & 8304.50 & 0.10\\
\hline $f_{10} - f_{00}$ & 1.224 & 4229.56 $\pm$ 0.29 & 4170.57 & 1.39\\
\hline $f_{11} - f_{01}$ & 1.224 & 4809.72 $\pm$ 0.30 & 4780.09 & 0.62\\
\hline $f_{01} - f_{00}$ & 1.224 & 7043.43 $\pm$ 0.43 & 7017.19 & 0.37\\
\hline $f_{11} - f_{10}$ & 1.224 & 7623.98 $\pm$ 0.44 & 7626.71 & 0.04\\
\hline 
\end{tabular}
\addtolength{\tabcolsep}{-5pt}
\end{center}
\end{table*}
\begin{table}
\begin{center}
\caption{\label{tab:cap_EM} Capacitance properties from electromagnetic simulations and design guide.}
\addtolength{\tabcolsep}{+5pt}
\begin{tabular}{
  |c|c| 
}
\hline 
Property & 
Value \\
\hline 
\hline
Transmon A's capacitance to ground & 61.9 fF \\
\hline 
Transmon B's capacitance to ground & 52.4 fF \\
\hline 
Capacitance between transmon A and B & 1.9 fF \\
\hline 
Josephson junction capacitance per area & $\text{57 fF}/\mu\text{m$^2$}$ \\
\hline 
\end{tabular}
\addtolength{\tabcolsep}{-5pt}
\end{center}
\end{table}
\begin{table}
\begin{center}
\caption{\label{tab:JJ_theory} JJ properties from fit results.}
\addtolength{\tabcolsep}{+5pt}
\begin{tabular}{
  |c|c| 
}
\hline 
Fit parameter & 
Fit result \\
\hline 
\hline
Critical current density $J_c$ & $\text{0.752 }  \mu\text{A}/\mu\text{m$^2$}$ \\
\hline 
Transmon A's JJ area& 0.026 $\mu\text{m$^2$}$\\
\hline 
Quarton's top JJ area& 0.077 $\mu\text{m$^2$}$\\
\hline 
Quarton's center JJ (left) area & 0.350 $\mu\text{m$^2$}$\\
\hline 
Quarton's center JJ (middle) area & 0.308 $\mu\text{m$^2$}$ \\
\hline 
Quarton's center JJ (right) area & 0.360 $\mu\text{m$^2$}$ \\
\hline 
Quarton's bottom JJ area & 0.077 $\mu\text{m$^2$}$ \\
\hline 
Transmon B's top JJ area & 0.037 $\mu\text{m$^2$}$ \\
\hline 
Transmon B's bottom JJ area & 0.049 $\mu\text{m$^2$}$ \\
\hline 
\end{tabular}
\addtolength{\tabcolsep}{-5pt}
\end{center}
\end{table}
Using the standard circuit quantization method, the device model is a function of the capacitances and Josephson energies of the JJs. Table \ref{tab:cap_EM} shows the capacitance properties from electromagnetic simulations and design guide. To estimate the Josephson energies of all the JJs in the device, we treat the junction areas and junction critical current $J_c$ as fit parameters to a cost function which takes into account the transition frequencies of Table~\ref{tab:freqs_vs_theory}, the self-Kerr of both qubits at $I_{\text{bias}}=1.285$ mA, and the two-tone spectroscopy of Fig.~\ref{fig: exp_2} at selected points. To numerically compute the energies of the system at different flux bias, we use the flux periods measured in section~\ref{section:flux-calibration}. We numerically diagonalize the Hamiltonian including internal modes, but neglecting the inductance through ground (see the potential from Eq.~(\ref{eq:full-potential})). We take scanning electron microscope (SEM) images of each JJ on a nominally identical device fabricated on the same wafer and use these measured areas as our initial guess (guessing also a critical current density $J_c = \text{0.7}  \mu\text{A}/\mu\text{m$^2$}$) to find a local minimum of the cost function. The JJ properties resulting from the fit are shown in Table~\ref{tab:JJ_theory}. The error with respect to the measured transition frequencies is shown in Table~\ref{tab:freqs_vs_theory}.

\section{\label{sec:transmon_freq_T1T2}Transmon transition frequencies and coherence times}
\begin{figure*}
    \includegraphics[width=0.75\textwidth]{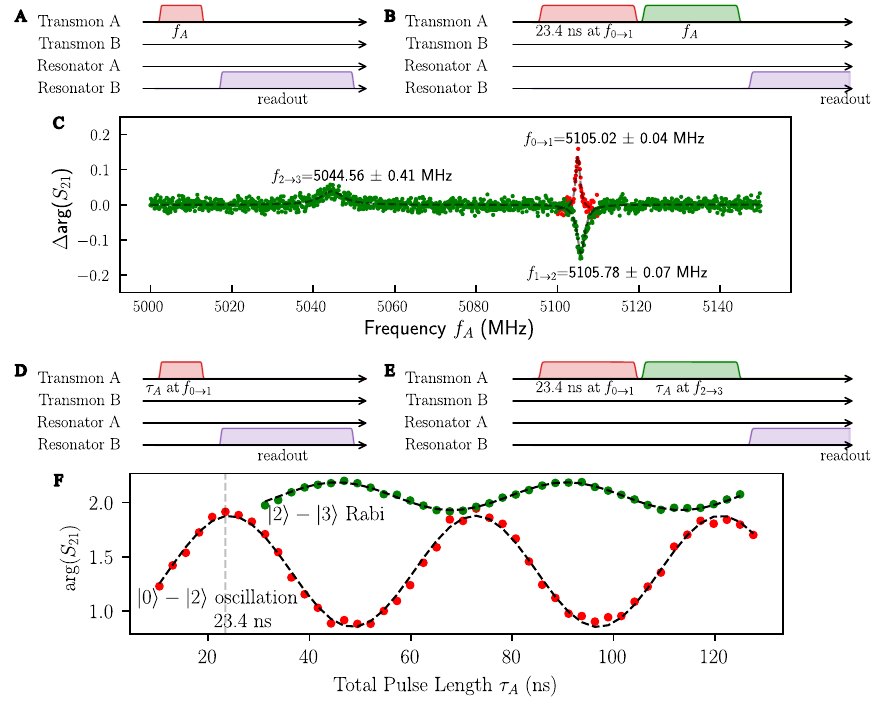}
    \caption{\textbf{Transmon A's transition frequencies at $I_{\text{bias}}=1.285$ mA.} (\textbf{A}) Pulse diagram for finding $f_{0\rightarrow1}$. (\textbf{B}) Pulse diagram for finding $f_{1\rightarrow2}$ and $f_{2\rightarrow3}$. (\textbf{C}) Readout results for $f_{0\rightarrow1}$ spectroscopy (red), and $f_{1\rightarrow2}$, $f_{2\rightarrow3}$ spectroscopy (green). Black dashed lines show Lorentzian fit. (\textbf{D}) Pulse diagram for driving $f_{0\rightarrow1}$ (and thus $\ket{0}-\ket{2}$ oscillation) over time $\tau_A$. (\textbf{E}) Pulse diagram for Rabi driving $f_{2\rightarrow3}$. (\textbf{F}) Readout results for $\ket{0}-\ket{2}$ oscillation (red), and $\ket{2}-\ket{3}$ Rabi oscillation (green), the latter is plotted with $f_{0\rightarrow1}$ pulse time included. Black dashed lines show cosine fit and grey dashed line shows $\tau_A = 23.4$ ns peak.}
    \label{fig: low_qb_anharm}
\end{figure*}

\begin{figure*}
    \includegraphics[width=0.85\textwidth]{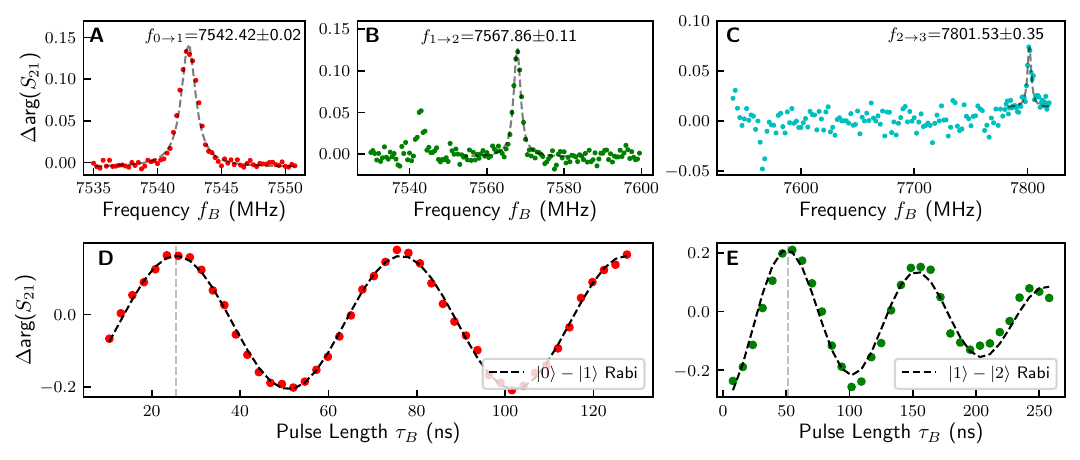}
    \caption{\textbf{Transmon B's transition frequencies at $I_{\text{bias}}=1.285$ mA.} (\textbf{A}) Spectroscopy results for $f_{0\rightarrow1}$. (\textbf{B}) Spectroscopy results for $f_{1\rightarrow2}$, obtained by taking spectroscopy after first applying $\pi$ pulse at $f_{0\rightarrow1}$. (\textbf{C}) Spectroscopy results for $f_{2\rightarrow3}$, obtained by taking spectroscopy after first applying $\pi$ pulse at $f_{0\rightarrow1}$ then $\pi$ pulse at $f_{1\rightarrow2}$. Black dashed lines in panels A-C show Lorentzian fit. (\textbf{D}) Rabi oscillation of $\{\ket{0},\ket{1}\}$ by driving at $f_{0\rightarrow1}$ with pulse length $\tau_B$, grey dashed line show calibrated $\pi$ pulse duration. (\textbf{E}) Rabi oscillation of $\ket{1,2}$ by first applying applying $\pi$ pulse at $f_{0\rightarrow1}$ then driving at $f_{1\rightarrow2}$ with pulse length $\tau_B$, grey dashed line show calibrated $\pi$ pulse duration. Black dashed lines in panels D-E show exponentially decaying cosine fit. }
    \label{fig: high_qb_anharm}
\end{figure*}

For quarton flux bias of $I_{\text{bias}}=1.285$ mA, Fig.~\ref{fig: low_qb_anharm}, Fig.~\ref{fig: high_qb_anharm} summarizes the measurement results for transition frequencies of transmon A, B, respectively. For the $f_{0\rightarrow1}$ transition, we perform standard pulsed two tone spectroscopy at very low drive amplitude to obtain a sharp transition peak. Then we observe Rabi (transmon B) or Rabi-like oscillations (linearized transmon A) in time by driving at $f_{0\rightarrow1}$, which enables calibration of $\pi$ pulse length (transmon B) or pulse length to approximately $\ket{2}$ (linearized transmon A). With this calibrated pulse applied first, a subsequent spectroscopy then reveals the higher level transitions, $f_{1\rightarrow2}$ (and $f_{2\rightarrow3}$ for linearized transmon A). We repeat this procedure for the higher level transitions, first calibrating $\pi$ pulse to $\ket{i}$ by observing Rabi oscillation, then performing spectroscopy for $f_{i\rightarrow(i+1)}$ after using previously calibrated pulses to drive the transmon to $\ket{i}$.


\begin{figure}
    \includegraphics[width=0.5\textwidth]{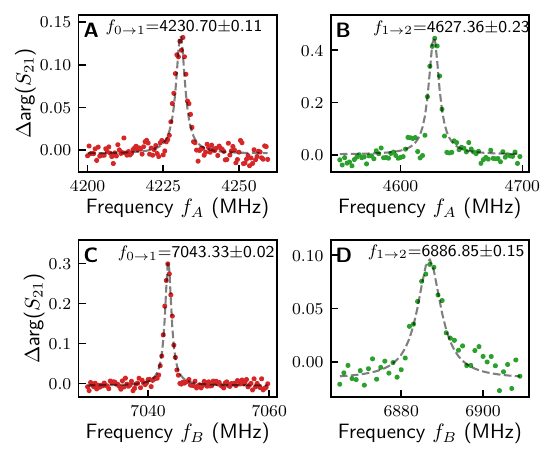}

    \caption{\textbf{Transmon A and B's transition frequencies at $I_{\text{bias}}=1.224$ mA.} (\textbf{A}) Spectroscopy results for transmon A's $f_{0\rightarrow1}$. (\textbf{B}) Spectroscopy results for transmon A's $f_{1\rightarrow2}$, obtained by taking spectroscopy after first applying $\pi$ pulse at $f_{0\rightarrow1}$.
    (\textbf{C}) Spectroscopy results for transmon B's $f_{0\rightarrow1}$. (\textbf{D}) Spectroscopy results for transmon B's $f_{1\rightarrow2}$, obtained by taking spectroscopy after first applying $\pi$ pulse at $f_{0\rightarrow1}$. Black dashed lines show Lorentzian fits. }
    \label{fig: 1.224_qb_anharm}
\end{figure}

We repeat this measurement protocol at the quarton flux bias of $I_{\text{bias}}=1.224$ mA, Fig.~\ref{fig: 1.224_qb_anharm} summarizes the measurement results for transition frequencies of both transmons. 

\begin{figure*}
    \includegraphics[width=0.85\textwidth]{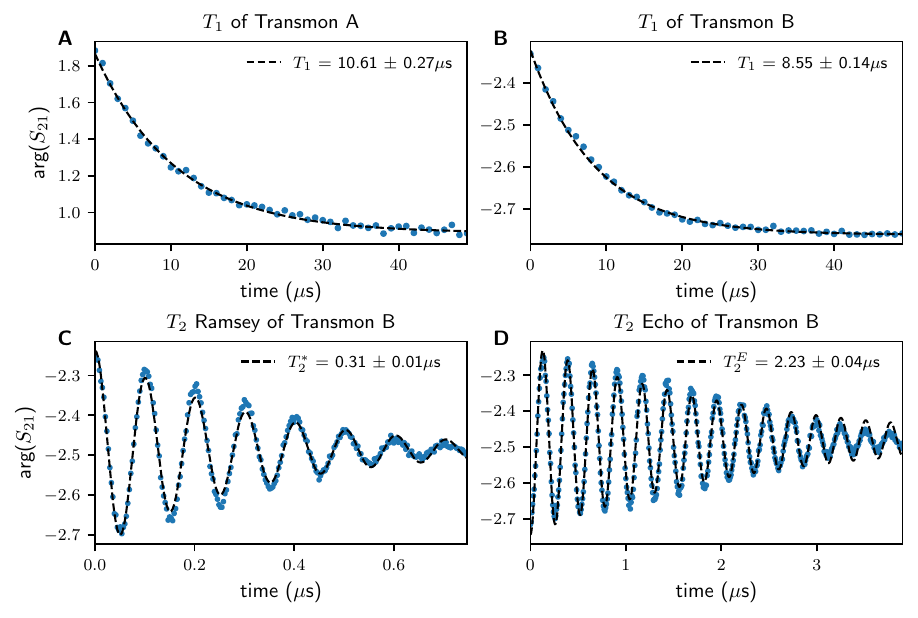}
    \caption{\textbf{Transmon coherence times at $I_{\text{bias}}=1.285$ mA.} 
    (\textbf{A}) $T_1$ of transmon A, measured by monitoring readout signal over time after first applying a 23.4 ns pulse at $f_{0\rightarrow1}$ (see Fig.~\ref{fig: low_qb_anharm}).
    (\textbf{B}) $T_1$ results of transmon B.
    (\textbf{C}) $T_2$ Ramsey results of transmon B.
    (\textbf{D}) $T_2$ echo results of transmon B. Black dashed lines show best fits with exponential decay (for $T_1$) or exponentially decaying cosine (for $T_2$) models.} 
    \label{fig:T1T2}
\end{figure*}

\begin{figure*}
    \includegraphics[width=0.85\textwidth]{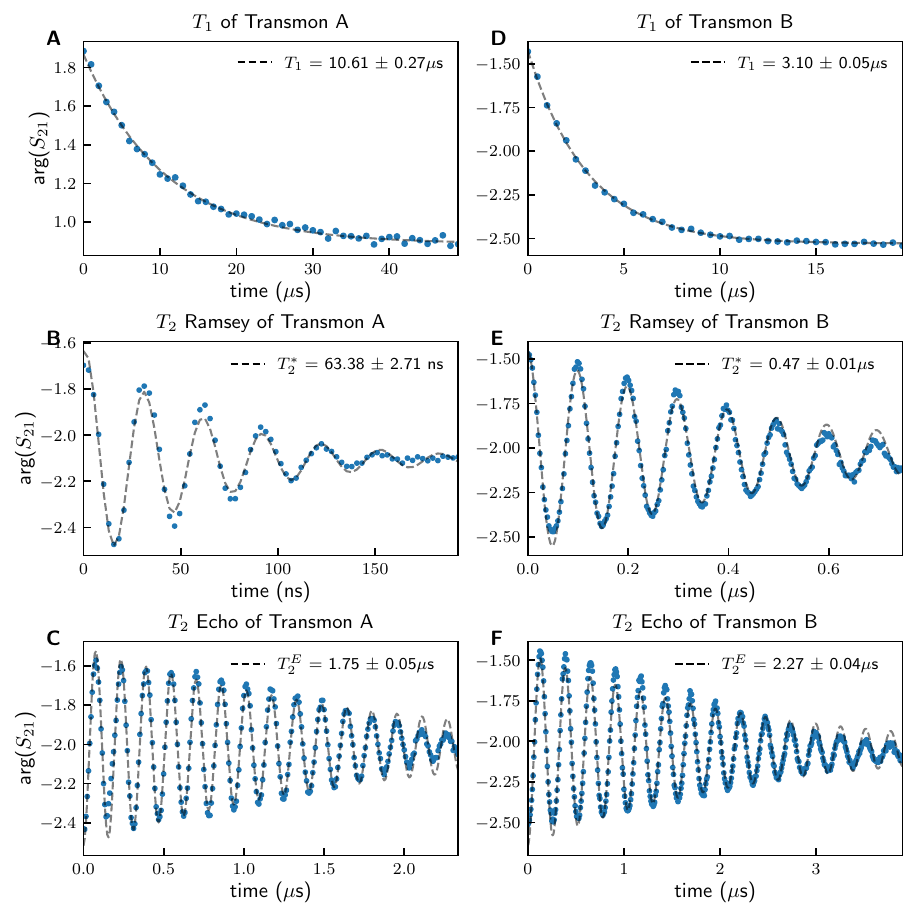}
    \caption{\textbf{Transmon coherence times at $I_{\text{bias}}=1.224$ mA.} 
    (\textbf{A-C}) $T_1$, $T_2$ Ramsey, and $T_2$ echo results of transmon A.
    (\textbf{D-F}) $T_1$, $T_2$ Ramsey, and $T_2$ echo results of transmon B. Black dashed lines show best fits with exponential decay (for $T_1$) or exponentially decaying cosine (for $T_2$) models.} 
    \label{fig:T1T2_1.224}
\end{figure*}

The coherence times of the transmons were measured with standard pulse sequences \cite{QICK}. Fig.~\ref{fig:T1T2} and Fig.~\ref{fig:T1T2_1.224} summarizes the measurement results at $I_{\text{bias}}=1.285$ and $I_{\text{bias}}=1.224$ mA, respectively. 

Table \ref{tab:1.224_param_table} summarizes the transition frequencies and coherence times of both transmons at $I_{\text{bias}}=1.224$ mA.
\vspace{0.45cm} 

\begin{table*}
\begin{center}
\caption{\label{tab:1.224_param_table}Summary of frequencies (MHz) and coherence times ($\mu$s) of both transmons at operating point where transmons have high anharmonicity ($I_{\text{bias}}=1.224$ mA).}
\addtolength{\tabcolsep}{+5pt}
\begin{tabular}{|c|c|c|c|c|c|}
\hline & $\chi/2\pi$ (MHz) & $f_{0\rightarrow1}$ (MHz) & $f_{1\rightarrow2} - f_{0\rightarrow1}$ (MHz) & $T_1$ ($\mu$s)  & $T_2^E$ ($\mu$s) \\
\hline \hline Transmon A & 580.54 $\pm$ 0.62 & 4230.70 $\pm$ 0.11 & 396.66 $\pm$ 0.25 & 10.61 $\pm$ 0.27 & 1.75 $\pm$ 0.05 \\
\hline Transmon B & 580.16 $\pm$ 0.42 & 7043.33 $\pm$ 0.02 & -156.47 $\pm$ 0.15 & 3.10 $\pm$ 0.05 & 2.27 $\pm$ 0.04 \\
\hline 
\end{tabular}
\addtolength{\tabcolsep}{-5pt}
\end{center}
\end{table*}

\section{\label{sec:LM-chi-extraction}Light-matter cross-Kerr extraction}
\begin{figure*}
    \includegraphics[width=1.0\textwidth]{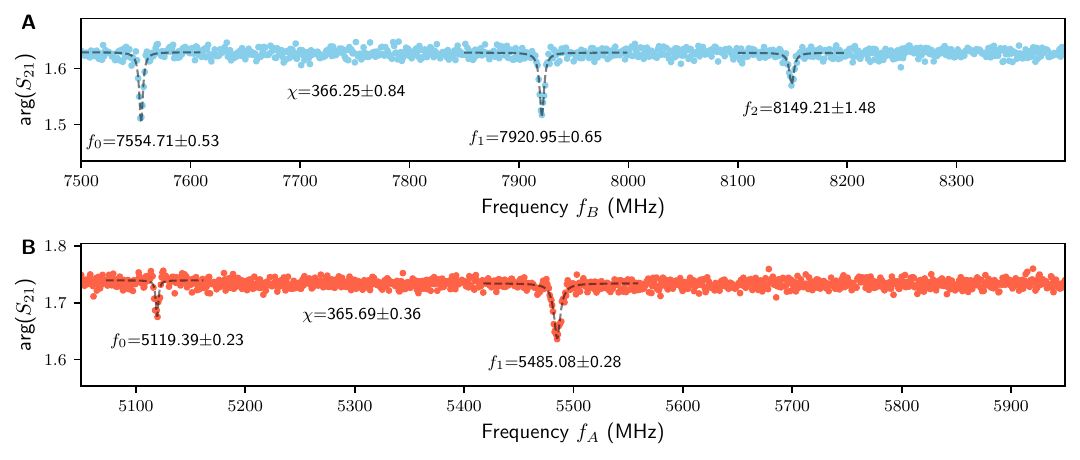}
    \caption{\textbf{Cross-Kerr values extracted from main text Fig.~\ref{fig: exp_3} measurements.} (\textbf{A}) Spectroscopy (c.f. main text Fig.~\ref{fig: exp_3}D) with Lorentzian fits to peaks (black dashed line). (\textbf{B}) Spectroscopy (c.f. main text Fig.~\ref{fig: exp_3}H) with Lorentzian fits to peaks (black dashed line). }
    \label{fig: cross-Kerr-extract}
\end{figure*}

Fig.~\ref{fig: cross-Kerr-extract} shows the cross-Kerr values extracted from the measurement described in Fig.~\ref{fig: exp_3} of main text. Note that compared to the main text, the data here uses a finer frequency step and lower probe power for more precise spectroscopy results. 

\section{\label{sec:light-light-in-time}Time response of simulated light-light nonlinear coupling}
\begin{figure*}
    \includegraphics[width=1\textwidth]{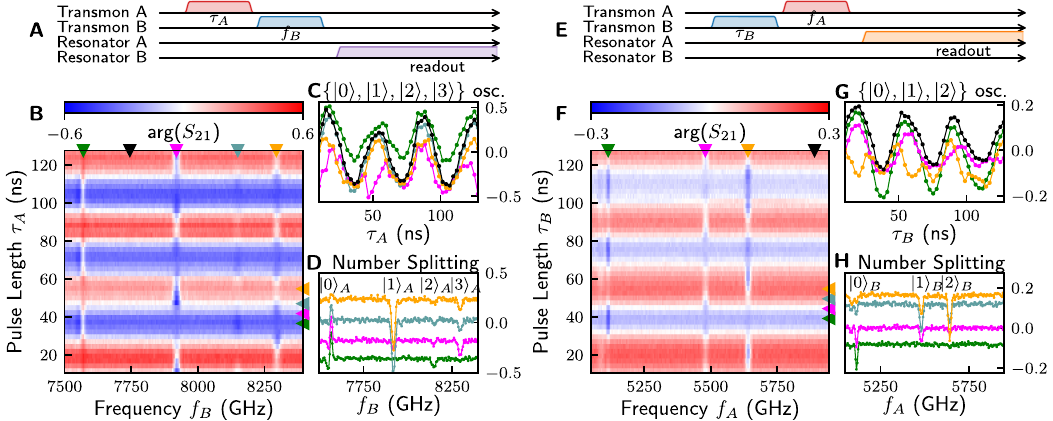}
    \caption{\textbf{Simulate light-light nonlinear coupling from detuned and strong drives.} (\textbf{A}) Pulse diagram:  detuned pulse of length $\tau_A$ to transmon A followed by pulse of frequency $f_B$ to transmon B. (\textbf{B}) Rabi-like oscillation and number splitting of transmon B spectrum by transmon A's excitation number $\{\ket{0},\ket{1},\ket{2}\}_A$. (\textbf{C}) Vertical line-cuts of panel B. (\textbf{D}) Horizontal line-cuts of panel B. (\textbf{E}) Pulse diagram:  detuned pulse of length $\tau_B$ to transmon qubit B followed by pulse of frequency $f_B$ to transmon A. (\textbf{F}) Rabi-like oscillation and number splitting of transmon A spectrum by transmon B's excitation number $\{\ket{0},\ket{1}\}_A$. (\textbf{G}) Vertical line-cuts of panel F. (\textbf{H}) Horizontal line-cuts of panel F.}
    \label{fig: detuned_crossKerr}
\end{figure*}

Fig.~\ref{fig: detuned_crossKerr} shows the results of simulated light-light nonlinear coupling with the same setup as in main text Fig.~\ref{fig: exp_3} but with stronger and frqeuency detuned first transmon drive. We observe Rabi-like oscillations but involving more higher energy states, similar to the response of linearized resonators. The system also shows more noise in time, which may be caused by the higher energy states having lower coherence times. Note also that data shown in Fig.~\ref{fig: detuned_crossKerr} uses coarser frequency step and higher spectroscopy probe power compared to the finer data shown in row 1 of main text Fig.~\ref{fig: exp_4}, 

\bibliography{scibib}

\end{document}